\documentclass[a4paper,aps,prd,onecolumn,preprintnumbers,showpacs,superscriptaddress,nofootinbib]{revtex4}
\usepackage{graphics,graphicx,epsfig}
\usepackage{amsfonts,amsmath,amssymb}

\begin{document}

\title{New Massive Gravity on de Sitter Space and Black Holes at the Special Point}

\author{Gregory Gabadadze}

\address{Center for Cosmology and Particle Physics, New York University\\
4 Washington Place, New York, NY 10003 USA}

\author{Gaston Giribet}

\address{Physics Department, University of Buenos Aires and IFIBA-CONICET\\
Ciudad Universitaria, Pabell\'on 1, 1428, Buenos Aires, Argentina}

\author{Alberto Iglesias}

\address{Center for Cosmology and Particle Physics, New York University\\
4 Washington Place, New York, NY 10003 USA}

\begin{abstract}
We study New Massive Gravity on de Sitter background and discuss several
interrelated properties: The appearance of an enhanced symmetry point at
linearized level where the theory becomes partially massless; its absence
at full nonlinear level, and its relation with the existence of static
black hole solutions and their hair parameter.
\end{abstract}


\maketitle

\section{Introduction}

A fully nonlinear covariant theory of massive gravity in $2+1$ dimensional spacetime, 
dubbed New Massive Gravity (NMG), has been introduced in Ref. \cite{Bergshoeff:2009hq}. At 
the linearized level, the theory is equivalent to the Fierz-Pauli action (FP) for a massive 
spin-2 field. It also passes highly nontrivial consistency checks at the nonlinear 
level\cite{alas, Gabadadzeetal}.

In this note, we revisit the formulation of NMG on de Sitter (dS) spacetime. We analyze the 
theory linearized about dS and study the appearance of an enhanced symmetry at a special 
point of the parameter space, similarly as it happens in FP gravity on dS 
\cite{Deser:1983mm,Higuchi:1986py,Deser:2001pe,Gabadadze:2008uc}. This symmetry enhancement 
has also been observed in Ref. \cite{Belgrado} within the canonical approach; here, we go 
further by relating this symmetry with the existence of dS black holes at the very special 
point of the parameter space.

\section{New massive gravity on de Sitter space}

The action of NMG is given by \cite{Bergshoeff:2009hq}
\begin{equation}
S=\frac{1}{16\pi G}\int d^3x~\sqrt{-g}
\left[R-2\lambda-\frac{1}{m^{2}}K\right]~, \ \ \ K=R_{\mu\nu}R^{\mu\nu}-\frac{3}{8}R^2~.   \label{action}
\end{equation}
The square-curvature terms $K$ satisfy the remarkable property $g^{\mu\nu}\delta K/\delta g^{\mu\nu}=K$, which is 
important for the 
unitarity of the theory about flat background.

The equations of motion derived from (\ref{action}) admit dS spacetime as exact solution provided $\lambda\leq m^2$, 
with the 
Hubble constant $H^2=2m^2(1\pm\sqrt{1-\lambda 
m^{-2}})$.

We consider perturbations of the form $g_{\mu\nu}=\gamma_{\mu\nu}+h_{\mu\nu}$, where 
$\gamma_{\mu\nu}$ denotes the metric of dS$_3$ space. We denote the covariant derivative 
with respect to the background metric by $\nabla$ and $\nabla_\mu\gamma^{\mu\nu}\nabla_\nu$ 
by $\Box$. It is possible to choose transverse traceless gauge $\nabla_\rho 
h^{\rho}_\mu=0,$ $h=\gamma^{\mu\nu}h_{\mu\nu}=0$, and the equations of motion read
\begin{equation}\label{eomtt}
-\frac{1}{2}\Box h_{\mu\nu}+\lambda h_{\mu\nu}
-\frac{1}{2m^2}\left(-\Box^2 h_{\mu\nu}+2H^2\Box h_{\mu\nu}
+\frac{5}{2}H^2\Box h_{\mu\nu}-\frac{11}{2}H^4h_{\mu\nu}\right)=0~,
\end{equation}
where we used that $ \nabla^\rho\nabla_\mu h_{\nu\rho}=\nabla_\mu\nabla^\rho h_{\nu\rho}
+3H^2h_{\mu\nu}-H^2\gamma_{\mu\nu}h$. Interestingly, the equations (\ref{eomtt}) factorize into
\begin{equation}
(\Box-m^2-\frac{5}{2}H^2)(\Box-2H^2)h_{\mu\nu}=0~.
\end{equation}
This splits the space of solutions into two: the General Relativity modes, $h_{\mu \nu}$, 
that solve 
equation $(\Box-2H^2)h_{\mu\nu}=0$, and the massive mode, $\tilde{h}_{\mu \nu}$, that 
solves $(\Box-m^2-5H^2/2)\tilde{h}_{\mu\nu}=0$. In the limit $m\to \infty$, the 
latter decouples.

\section{Conformal symmetry at the special 
point}

Here, we will be concerned with the special points $\lambda=m^2$, that is $H^2=2m^2$. At this point, the theory 
exhibits a special 
property. To see this, consider the variation of action (\ref{action}) under the transformation 
\begin{equation}
\delta g_{\mu\nu}=\gamma_{\mu\nu}\phi~,  \label{delg}
\end{equation}
for an arbitrary function $\phi$. Up to a total derivative, the variation of the action is
\begin{equation}
\delta S=\int d^3x \sqrt{-g}\left[\frac{1}{2}\left(1-\frac{H^2}{2m^2}\right)
(\Box h -\nabla_\rho \nabla_\sigma h^{\rho\sigma})
+\left(\lambda-\frac{H^4}{4m^2}\right)h\right]\phi~.
\end{equation}

We see that, if $\lambda=m^2$ (equivalently, $H^2=2m^2$), the variation does vanish for 
arbitrary $\phi$; therefore, at this point the linarized theory exhibits a symmetry 
enhancement. This is reminiscent of the special point of FP gravity on dS. In the case of 
FP theory, the enhanced symmetry is of the form $ \delta 
g_{\mu\nu}=(H^{-2}\nabla_\mu\nabla_\nu+ \gamma_{\mu\nu})\phi$. NMG 
is generally covariant, so that the term $\nabla_\mu \nabla_\nu\phi$ represents 
a 
symmetry by itself; it is the $\gamma_{\mu\nu}\phi$ term that becomes enhanced at 
$\lambda=m^2$.

\section{Black holes at the special point}

It turns out that, at $\lambda=m^2$, NMG exhibits another peculiar property: it admits dS 
black hole solutions 
\cite{Bergshoeff:2009aq, Oliva:2009ip}. The metrics is
\begin{equation}
ds^2=-\left(- (Hr)^2+2bHr-\mu\right)dt^2+\frac{dr^2}{-
(Hr)^2+2bHr-\mu}+r^2 d\theta^2, \label{bhmetric}
\end{equation}
where $b$ and $\mu $ are two arbitrary parameters. In the range $0<\mu\le b^2$, metric 
(\ref{bhmetric}) describes a black hole that asymptotes dS$_3$ spacetime. The black hole 
exhibits a curvature 
singularity at the origin, it being covered by an event horizon located at 
$r_-=(b/H)(1-\sqrt{1-\mu /b^2})$. The cosmological horizon of dS$_3$ space is located at 
$r_+=(b/H)(1+\sqrt{1-\mu /b^2})$.

Considering $b$ as small parameter and expanding to linear order one gets 
\begin{equation} 
ds^2\approx g^0_{\mu\nu} dx^{\mu} dx^{\nu} - 2bHr \ dt^2- 
\frac{2bHr}{(\mu+(Hr)^2)^2}dr^2, \label{bhlin} 
\end{equation} 
where $g^0_{\mu\nu}$ 
corresponds to metric (\ref{bhlin}) with $b=0$. Now, we observe that a special combination 
of a conformal symmetry and coordinate transformations of the form 
\begin{equation}\label{trans} 
\delta g_{\mu\nu}=\nabla_{(\mu}\xi_{\nu)}+g^0_{\mu\nu}\phi, \ 
\ {\text with } \ \ \xi^\nu=-\frac{bHr^2}{\mu}\delta^\nu_r~, \ \ \ \ 
\phi=\frac{2bHr}{\mu}~, 
\end{equation} 
precisely cancels the terms proportional to $b$ in 
(\ref{bhlin}). This means that (\ref{bhlin}) can be generated by a transformation 
(\ref{delg}). At the nonlinear level, the symmetry (\ref{delg}) is absent.

Away from the special point, the linear analysis of NMG
reveals the existence of a non-propagating tensor mode of 3D 
massless General Relativity, as well as the helicity-$2$, 
helicity-$1$, and the helicity-$0$ modes of a 3D massive graviton
(helicity-$2$ being nondynamical). In contrast, at the special point, 
the kinetic term of the helicity-$0$ mode vanishes, and it 
becomes infinitely strongly coupled since the corresponding conformal 
symmetry is present at the linear level only.
These facts get  reflected onto  the black hole solution (6) 
as follows:
as shown above, the black hole  hair is a gauge artifact in the 
linearized theory, while this hair is not removable in the full 
nonlinear case\footnote{Note, however, that on the full 
black hole  solution (6) the helicity-$0$ model may or may not 
acquire its kinetic term due to nonzero $\mu$ and/or $b$.}.

It appears that the consistency with no-hair theorems prevents the 
existence of the black holes away from the special point since 
for $\lambda \neq m^2$  the extra longitudinal
mode, which would provide the hair, is a propagating 
field.  At $\lambda= m^2$  the black holes are 
possible only in the regime where the hair is carried
by the longitudinal mode that is very (or perhaps infinitely) 
strongly coupled, at least when the dS horizon is approached. 
This feature is what seems to be responsible for 
the evasion of the no-hair theorems.
It may be interesting to study similar questions  
in 3D ghost-free massive gravity \cite{deRham:2010ik,deRham:2010kj}. 
The latter theory differs from NMG by the absence in it of 
the massless 3D GR field. This field is responsible for negative mass, $-\mu$ in (\ref{bhmetric}).

\end{document}